\documentclass[12pt]{elsarticle}
\usepackage{graphicx}
\usepackage{epsfig}
\usepackage{amsmath}
\usepackage{graphics}
\usepackage{epsfig}
\begin{document}

\title{Restoring the Pauli principle in the random phase approximation  ground state}

\author{D. S. Kosov }
\address{ College of Science and Engineering, James Cook University, Townsville, QLD, 4811, Australia}

\begin{abstract} 
Random phase approximation ground state contains electronic configurations where two (and more) identical electrons can occupy the same molecular spin-orbital violating the Pauli exclusion principle.
This overcounting of electronic configurations happens due to quasiboson approximation in the treatment of electron-hole pair  operators. We describe the method to restore the Pauli principle in the RPA wavefunction. The proposed theory is illustrated by the calculations of molecular dipole moments and electronic kinetic energies. The Hartree-Fock based RPA, which is corrected for the Pauli principle, gives the results of comparable accuracy with  M{\o}ller-Plesset second order perturbation theory  and coupled-cluster singles and doubles method.
\end{abstract}
\maketitle

\newpage
\section{Introduction}
 Random phase approximation  (RPA)  has recently emerged as a  very promising method to study correlated ground states in molecules \cite{Eshuis2012,yang13,goerling13,rocco14,mussard15,kallay15,colonna16,scuseria15}.
 RPA calculations, especially when they are performed based on the density functional theory,  give  accurate total molecular 
 correlation energies, reaction barriers, electron affinities, ionisation potentials \cite{Eshuis2012,yang13,goerling13,rocco14,mussard15,kallay15,colonna16,scuseria15,furche10,schurkus16,scuseria12,rekkedal,furche14} 
 and provide  a correct description of van der Waals intermolecular interactions \cite{dobson12,PhysRevLett.102.096404,tkatchenko15}. The connection of RPA with the coupled-cluster doubles approach has been recently established \cite{yang13,scuseria08},
 which puts the method into the context of traditional quantum chemistry practices.

 In this paper we would like to draw  attention to the problem of violation of the Pauli exclusion principle in the RPA ground state and discuss the way of correcting it.
This problem has been  known in nuclear physics since 1968 \cite{rowe68} and since then,  has been a subject of intensive studies there \cite{ELLIS1970625,Lenske1990,mishev13}. In quantum chemistry,  despite early pioneering works of Simons  \cite{simons71},  Linderberg and Ohrn  \cite{linderberg77}, the 
 issues associated with the antisymmetry of RPA wavefunction
have been largely overlooked.
 To understand the problem, let us look at the (schematically written) RPA ground state wavefunction \cite{rowe,ring_and_schuck}
 \begin{equation}
 |\Psi_0 \rangle \sim e^{\sum T_{p_1 h_1 p_2 h_2} c^\dag_{p_1 } c_{h_1} c^\dag_{p_2} c_{h_2} } |HF\rangle.
 \label{rpa-psi}
 \end{equation}
Here and throughout the paper, indices $h$ and $p$ refer to Hartree-Fock occupied and virtual molecular orbitals, respectively, and $c^\dag_{k} (c_{k})$  creates (annihilates) an electron in  the Hartree-Fock molecular orbital $k$. $|HF \rangle $ is the Hartree-Fock ground state vector. The spin index is suppressed in the introduction for the simplicity. RPA treats particle-hole pair creation operators $c^\dag_p c_h$ as  bosons, approximating the commutator (so-called quasiboson approximation\cite{rowe,ring_and_schuck}) as
\begin{equation}
[c^\dag_{h_2} c_{p_2},c^\dag_{p_1} c_{h_1} ] \approx \delta_{p_1 p_2} \delta_{h_1 h_2},
\label{qba}
\end{equation}
 and therefore neglecting the internal fermionic structure of these operators. This leads to the over-counting of  configurations in the RPA ground state wavefunction. The configurations  with $p_1 = p_2$ and $h_1=h_2$ in the exponent (\ref{rpa-psi}), which otherwise would be forbidden by the Pauli principle, do contribute to  the RPA ground state.
 The violation of the Pauli principle is amplified in the RPA due to the exponential structure of the ground state -- expanding the exponent we see that  terms containing two, four, six, eight and more identical electrons are allowed to occupy the same  single-particle state. 
Therefore, the RPA ground state wavefunction (if it is treated in quasiboson approximation) over-correlates the electronic systems by including unphysical configurations  and by letting the electrons to come closer in the violation of the Pauli  principle.
 
 How can we remove these unphysical states from the RPA ground state wavefunction and restore the Pauli principle? How can we correct  the final expression for computed observables  for the Pauli principle and at the same time preserve the original RPA equations? These  are the questions which interest us in this paper.

\section{Theory}
\subsection{RPA ground state vector with the Pauli principle corrections}

We will work with the  Hartree-Fock quasiparticle creation and annihilation operators,  $a^\dag_{k\sigma}$ and $a_{k\sigma}$, which are related to the original electron creation and annihilation operators in the Hartree-Fock spin-orbital,  $c^\dag_{k\sigma}$ and $c_{k\sigma}$, via 
the transformations \cite{surjan}:
\begin{eqnarray}
\label{a+}
a^\dag_{k\sigma} =\left\{   
\begin{array}{l}
c^\dag_{k\sigma}, \text{   if } k \text{ is virtual orbital},
\\
c_{k\sigma}, \text{   if } k \text{ is occupied orbital},
\end{array}
\right.
\end{eqnarray} 
and
 \begin{eqnarray}
 \label{a}
a_{k\sigma} =\left\{   
\begin{array}{l}
c_{k\sigma}, \text{   if } k \text{ is virtual orbital},
\\
c^\dag_{k \sigma}, \text{   if } k \text{ is occupied orbital}.
\end{array}
\right.
\end{eqnarray} 
Here  $\sigma=-1/2, 1/2$  are two projections of  electron spin.
The  quasiparticle operators defined above have Hartree-Fock ground state $|HF\rangle$ as a vacuum
\begin{equation}
a_{k\sigma}  |HF\rangle =0 \text{      for all  } k, \sigma.
\end{equation}

The building blocks for the RPA wavefunction are the particle-hole pair creation operators  $C^{\dagger}_{ph}(JM)$ defined as
\begin{equation}
C^{\dagger}_{ph}(JM) = \sum_{\sigma \sigma'} \langle\frac{1}{2} \sigma \frac{1}{2} \sigma' | JM \rangle a^\dagger_{p\sigma}  a^\dagger_{h \overline{\sigma'}}.
\end{equation}
It creates the  particle-hole excited pair with  spin $J$ and spin projection $M$.
Over-line over spin indices means  angular momentum time-reversal state  $C_{ph}(\overline{JM}) = (-1)^{J+M}C_{ph}({J-M}) $ and $a^\dagger_{h \overline{\sigma}}= (-1)^{1/2+\sigma} a^\dagger_{h-\sigma}$. 
The Clebsch-Gordan coefficient $\langle\frac{1}{2} \sigma \frac{1}{2} \sigma' | JM \rangle$ couples electronic spins of occupied and virtual molecular orbitals into singlet $J=0, M=0$ or triplet  $J=1, M=-1,0,1$ states.

The RPA ground state wavefunction has the following form\cite{rowe,ring_and_schuck}
\begin{equation}
|\Psi_0 \rangle = N e^S |HF\rangle,
\label{psi0}
\end{equation}
where
\begin{equation}
S= \frac{1}{2} \sum_{JM} \sum_{php'h'} T^J_{php'h'} C^\dag_{ph}(JM)  C^\dag_{p'h'}(\overline{JM}).
\label{S}
\end{equation}
Here and everywhere in the paper
\begin{equation}
\sum_{JM} = \sum_{J=0}^1 \sum_{M=-J}^J .
\end{equation}
The correlated  RPA ground state  $|\Psi_0  \rangle $ is a vacuum for the RPA excitation annihilation operators
\begin{equation}
Q_{JMi}  |\Psi_0\rangle =0,  \text{      for all  } J, M, \text{ and } i,
\label{vacuum}
\end{equation}
where  the  excitation annihilation operator is obtained from the corresponding creation operator
 \begin{equation}
Q^\dag_{JMi} = \sum_{ph} X_{ph}^{Ji} C^\dag_{ph}(JM) - Y_{ph}^{Ji} C_{ph}(\overline{JM})
\label{Q+}
\end{equation}
via Hermitian conjugation.

The  structure of operator $Q^\dag_{JMi}$ and related  RPA  excitation energy  $\omega_{Ji}$
are determined from  the standard system of RPA equations, which we write   separately for singlet ($J=0$) and triplet ($J=1$) electronic excitations
\begin{eqnarray}
\label{rpa1}
\sum_{p'h'} A^J_{ph,p'h'} X^{Ji}_{p'h'} + B^J_{ph,p'h'} Y^{Ji}_{p'h'} = \omega_{Ji} X^{Ji}_{ph}, \\
\label{rpa2}
\sum_{p'h'} B^J_{ph,p'h'} X^{Ji}_{p'h'} + A^J_{ph,p'h'} Y^{Ji}_{p'h'} = -\omega_{Ji} Y^{Ji}_{ph}.
\end{eqnarray}
Matrices $A$ and $B$ are given by the following expressions
\begin{eqnarray}
A^J_{ph,p'h'} = (\epsilon_p-\epsilon_h) \delta_{pp'} \delta_{hh'} 
 +\left[ 1+(-1)^J \right] (ph|p'h') -(hh'|p'p),
\end{eqnarray}
\begin{equation}
B^J_{ph,p'h'} =  \left[ 1+(-1)^J \right] (ph|p'h') -(p'h|ph'),
\end{equation}
\\
where  $\epsilon_k$ is Hartree-Fock energy of $k$ molecular orbital  and $(kl|mn)$ is a two-electron integral in the Mulliken notations. 

Suppose that  we know $X^{Ji}_{ph}$ and $Y^{Ji}_{ph}$ in (\ref{Q+}) from the solution of the RPA equations.
Our goal in this section is to solve  (\ref{vacuum})  directly and determine coefficient $T$  without the use of the quasiboson approximation (\ref{qba}).
We rewrite (\ref{vacuum}) as
\begin{equation}
Q_{JMi} e^S |HF \rangle =0
\end{equation}
Using the standard operator identity we get
\begin{eqnarray}
e^S ( Q_{JMi}  + [Q_{JMi} ,S] +
\frac{1}{2!} [[Q_{JMi},S],S] + ... )  |HF \rangle =0.
\end{eqnarray}
We neglect higher order terms  $O(S^2)$ in this expansion and the equation for operator $S$ becomes
\begin{equation}
 ( Q_{JMi}  + [Q_{JMi} ,S] ) |HF \rangle =0.
 \label{qqs}
\end{equation}
Substituting explicit expression for $Q_{JMi}$ into (\ref{qqs}) and rearranging the terms, we get

\begin{eqnarray}
\nonumber
\frac{1}{2} \sum_{JM} \sum_{p_1h_1p_2 h_2} \sum_{ p_3 h_3} T^J_{p_1 h_1p_2 h_2} X^{J_3 i}_{p_3 h_3} 
 \left[C_{p_3 h_3}(J_3 M_3), \; C^\dag_{p_1 h_1}(JM)  C^\dag_{p_2 h_2}(\overline{JM}) \right]  |HF \rangle 
 \\
= \sum_{p_3 h_3} Y^{J_3i}_{p_3 h_3} C^\dag_{p_3 h_3}( \overline{J_3 M_3}) |HF \rangle.
\label{qqs1}
\end{eqnarray}
The commutator in the l.h.s. of (\ref{qqs1}) can be exactly computed and (\ref{qqs1}) becomes
\begin{eqnarray}
 && \sum_{JM} \sum_{p_1h_1p_2 h_2} \sum_{ p_3 h_3} T^J_{p_1 h_1p_2 h_2} X^{J_3 i}_{p_3 h_3}   \Big\{ \; \delta_{JJ_3} \delta_{MM_3} \delta_{p_1 p_3} \delta_{h_1 h_3} 
 \nonumber
 \\
&& -
\frac{1}{2}   \sum_{\sigma_3 \sigma_3'  \sigma_1 \sigma_1'}  \langle \frac{1}{2} \sigma_3 \frac{1}{2} \sigma_3 |J_3 M_3  \rangle  \langle \frac{1}{2} \sigma_1 \frac{1}{2} \sigma_1' |J  M  \rangle 
\Delta(p_3\sigma_3, p_1 \sigma_1, h_3 \sigma_3', h_1 \sigma_1') 
\Big\} \; 
\nonumber
 \\
&&\times C^\dag_{p_2 h_2}(\overline{JM}) |HF \rangle 
=\sum_{p_3 h_3} Y^{J_3i}_{p_3 h_3} C^\dag_{p_3 h_3}( \overline{J_3 M_3}) |HF \rangle,
\label{qqs2}
\end{eqnarray}
where 
\begin{equation}
\Delta(p_1\sigma_1, p_2 \sigma_2, h_1 \sigma_1', h_2 \sigma_2') =    \delta_{p_1 p_2} \delta_{\sigma_1 \sigma_2}  
  a^\dag_{h_2 \overline{ \sigma_2' } } a_{h_1 \overline{ \sigma_1' } }  + \delta_{h_1 h_2} \delta_{\sigma_1' \sigma_2'}  
  a^\dag_{p_2   \sigma_2  } a_{p_1 \sigma_1}.
\end{equation}
To perform this derivation we use  exact commutator between particle-hole pair creation and annihilation operators
 \begin{eqnarray}
 \nonumber
 \left[ C_{p_1 h_1}(J_1 M_1), C^{\dagger}_{p_2h_2}(J_2 M_2)   \right]= \delta_{p_1p_2} \delta_{h_1 h_2} \delta_{J_1 J_2} \delta_{M_1 M_2}
 \\
  - \underbrace{\sum_{\sigma_1 \sigma_1'  \sigma_2 \sigma_2'}  \langle \frac{1}{2} \sigma_1 \frac{1}{2} \sigma_1' |J_1 M_1  \rangle  \langle \frac{1}{2} \sigma_2 \frac{1}{2} \sigma_2' |J_2 M_2  \rangle 
  \Delta(p_1\sigma_1, p_2 \sigma_2, h_1 \sigma_1', h_2 \sigma_2') }_{\text{Pauli correction}}.
  \label{commutator}
 \end{eqnarray} 
 The commutator (\ref{commutator}) deserves  special discussion. If we kept only the first term in r.h.s (\ref{commutator}), this would correspond to standard  RPA quasiboson approximation. The Pauli correction part is roughly proportional to the  total number of particle and hole excitations in the RPA state, that means the more  RPA ground state deviates from  Hartree-Fock Stater determinant, the larger the Pauli corrections contribution  to the r.h.s of the commutator (\ref{commutator}) become and the more pronounced deviation from the bosonic behaviour of operators $C^\dag(C)$ we see.
 
Coming back to (\ref{qqs2}) and taking into account that $|HF\rangle$ is vacuum for the Hartree-Fock quasiparticle operators, after some straightforward algebra we transform (\ref{qqs2}) to the following form:
\begin{equation}
\sum_{p'h'} T^J_{php'h'} X^{Ji}_{p'h'} -  \underbrace{\sum_{p'h' J'} T^{J'}_{ph'p'h} W(J',J) X^{Ji}_{p'h'}}_{\text{Pauli correction}} =Y^{Ji}_{ph}.
\label{tx}
\end{equation}
Here matrix $W$ is
\begin{eqnarray}
  \label{w}
&&W(J',J) =  \sum_{M'=-J'}^{J'} \sum_{\sigma_1 \sigma_1'  \sigma_2 \sigma_2'}  (-1)^{J+J'+M+M'}
\\
 &&\times \langle \frac{1}{2} \sigma_2 \frac{1}{2} \sigma_1' |J M  \rangle  
 \langle \frac{1}{2} \sigma_1 \frac{1}{2} \sigma_1' |J'  M'  \rangle 
 \langle \frac{1}{2} \sigma_2 \frac{1}{2} \sigma_2' |J' -M'  \rangle 
   \langle \frac{1}{2} \sigma_1 \frac{1}{2} \sigma_2' |J  -M  \rangle.
   \nonumber
\end{eqnarray}
Matrix $W$ (\ref{w}) is calculated numerically by summing up the corresponding Clebsch-Gordan coefficients and it is
 \begin{eqnarray}
 W(0,0) =\frac{1}{2}, \;\;\; W(0,1) =- \frac{1}{2}, 
 \\
 W(1,0) =- \frac{3}{2},  \;\;\;  W(1,1) =-\frac{1}{2}.
 \end{eqnarray}
The similar to (\ref{tx}) equation was obtained to describe RPA  ground state correlations in spherical nuclei  by Lenske and Wambach \cite{Lenske1990}.
 \subsection{Calculation single-particle density matrices}
 
Many molecular observables can be expressed in terms of single-particle density matrix.
We define single-particle density matrix operators as
\begin{equation}
\rho_{p'p} = \sum_\sigma a^\dag_{p \sigma} a_{p' \sigma},
\;\;\;\;
\rho_{h'h} = \sum_\sigma a^\dag_{p \sigma} a_{p' \sigma}.
\end{equation}
We show detailed derivations of the expectation value for operator $\rho_{p'p}$ in the RPA ground state. The calculations for $\rho_{h'h}$ are almost identical, we simply write the final expression for its expectation value in the end of the derivations.

We begin with the calculations of two commutators
\begin{equation}
[\rho_{p'p}, C_{p_1 h_1}^\dag(JM)] = \delta_{p'p_1} C^\dag_{p h_1} (JM),
\end{equation}

\begin{equation}
[\rho_{p'p}, S] = 
\sum_{JM} \sum_{h_1 p_2 h_2} T^J_{p' h_1 p_2 h_2} C^\dag_{p h_1} (JM)  
C^\dag_{p_2 h_2} (\overline{JM}).
\label{rhoS}
\end{equation}
The expectation value is
\begin{eqnarray}
\langle \Psi_0 |\rho_{p'p} | \Psi_0 \rangle = \langle \Psi_0 | \rho_{p'p}  e^S |HF \rangle =
\nonumber
\\
\langle \Psi_0 |e^S\left( \rho_{p'p} +[\rho_{p'p},S] + \frac{1}{2} [[\rho_{p'p},S],S]+ ...  \right) |HF \rangle.
\nonumber
\end{eqnarray}
Taking into account that  $\rho_{p'p} |HF \rangle =0$ (since $|HF\rangle$ is vacuum for Hartree-Fock annihilation operators) and that $[[\rho_{p'p},S],S] =0$  (since $[\rho_{p'p},S]\sim C^\dag $) so that  it and all higher terms vanish, we get
\begin{equation}
\langle \Psi_0 |\rho_{p'p} |\Psi_0 \rangle =\langle \Psi_0 |e^S [\rho_{p'p},S] |HF \rangle.
\end{equation}
 Then, again, since the commutator $[\rho_{p'p},S] $ commutes with $S$, the expectation value becomes
\begin{equation}
\langle \Psi_0 |\rho_{p'p} | \Psi_0 \rangle =\langle \Psi_0 | [\rho_{p'p},S] |\Psi_0 \rangle.
\label{avr}
\end{equation}
Substituting the commutator (\ref{rhoS})  into (\ref{avr}) we get
\begin{eqnarray}
\label{avr1}
 \langle \Psi_0 |\rho_{p'p} |\Psi_0 \rangle =
\sum_{JM} \sum_{h_1 p_2 h_2} T^J_{p' h_1 p_2 h_2}  \langle \Psi_0 | C^\dag_{p h_1} (JM)  
C^\dag_{p_2 h_2} (\overline{JM}) |\Psi_0 \rangle.
\end{eqnarray}
We express  $C^\dag_{p_2h_2}(\overline{JM}) $ in terms of RPA excitation  and annihilation operators 
 \begin{equation}
 C^\dag_{p_2h_2}(\overline{JM}) = \sum_i X^{Ji}_{p_2 h_2} Q^\dag_{\overline{JM}i} + Y^{Ji}_{p_2 h_2} Q^\dag_{{JM}i}
 \end{equation}
and substitute it back into  equation (\ref{avr1}):
 
 \begin{eqnarray}
 \label{avr2}
\langle \Psi_0 |\rho_{p'p} | \Psi_0 \rangle 
 = \sum_{i}  \sum_{JM} \sum_{h_1 p_2 h_2} T^J_{p' h_1 p_2 h_2} X^{Ji}_{p_2 h_2} \langle \Psi_0 | C^\dag_{p h_1} (JM)  
Q^\dag_{\overline{JM} i} |\Psi_0 \rangle.
\nonumber
\end{eqnarray}
 Contracting the amplitude $X$ and coefficient $T $ with the use of equation (\ref{tx}), we get
  \begin{eqnarray}
  \nonumber
\langle \Psi_0 |\rho_{p'p} | \Psi_0 \rangle =
\sum_{i}  \sum_{JM} \sum_{h_1 }
\left(Y^{Ji}_{p' h_1} + \sum_{p_3h_3 J_1} T^{J_1}_{p' h_3 p_3 h_1} W(J_1,J) X^{Ji}_{p_3 h_3} \right)
\\
\times
 \langle \Psi_0 | C^\dag_{p h_1} (JM)  
Q^\dag_{\overline{JM} i} |\Psi_0 \rangle.
\label{avr3}
\end{eqnarray}
 
The calculation of the first term in (\ref{avr3}) is straightforward
  \begin{eqnarray}
\label{first}  
&& \sum_{i} \sum_{JM}  \sum_{h_1 }
Y^{Ji}_{p' h_1}  \langle \Psi_0 | C^\dag_{p h_1} (JM)  
Q^\dag_{\overline{JM} i} |\Psi_0 \rangle
\\
&&
= 
 \sum_{i i'} \sum_{JM}  \sum_{h_1 }
Y^{Ji}_{p' h_1}  Y^{Ji'}_{p h_1}  \underbrace{ \langle \Psi_0 | Q_{\overline{JM} i'}
Q^\dag_{\overline{JM} i} |\Psi_0 \rangle}_{\delta_{ii'}}
= \sum_{i} \sum_{JM}   \sum_{h_1 }
Y^{Ji}_{p' h_1}  Y^{Ji}_{p h_1} 
\nonumber
\end{eqnarray}
The calculation of the second term in (\ref{avr3}) is more involved and is given in some details in the appendix. The result is very interesting: The second term in (\ref{avr3}), that is the Pauli principle corrections to the expectation value of single particle density matrix $\rho_{p'p}$, is the expectation value of the density matrix itself taken with negative sign:
  
  \begin{eqnarray}
  \nonumber
\sum_{J_1}  \sum_{i} \sum_{JM}
 \sum_{h_1 p_2 h_2 } T^{J_1}_{p' h_2 p_2 h_1} 
 W(J_1,J) X^{Ji}_{p_2 h_2} 
 \langle \Psi_0 | C^\dag_{p h_1} (JM)  
Q^\dag_{\overline{JM} i} |\Psi_0 \rangle 
\\
= - \langle \Psi_0 | \rho_{p'p} |\Psi_0 \rangle
\label{second}
\end{eqnarray}

 Combining these two terms (\ref{first},\ref{second}) together into (\ref{avr2})  we get
 \begin{equation}
\langle \Psi_0 |\rho_{p'p} |\Psi_0 \rangle =   \sum_{i} \sum_{JM} \sum_{h }
Y^{Ji}_{p' h}  Y^{Ji}_{p h}  - \underbrace{ \langle \Psi_0 |\rho_{p'p} |\Psi_0 \rangle}_{\text{Pauli correction}},
\label{rhopp-P}
 \end{equation}
 which means
  \begin{equation}
\langle \Psi_0 |\rho_{p'p} |\Psi_0 \rangle = \underbrace{\frac{1}{2}}_{\text{Pauli correction}}  \sum_{i} \sum_{JM}  \sum_{h }
Y^{Ji}_{p' h}  Y^{Ji}_{p h}.
\label{rhopp-P1}
 \end{equation}
 The similar calculations are performed for single-particle density matrix for occupied molecular orbitals
 with the result
   \begin{equation}
\langle \Psi_0 |\rho_{h'h} |\Psi_0 \rangle = \underbrace{\frac{1}{2}}_{\text{Pauli correction}}  \sum_{i} \sum_{JM} \sum_{p }
Y^{Ji}_{p h'}  Y^{Ji}_{p h}.
\label{rhohh-P1}
 \end{equation}

\section{Test results}

We  illustrate the  role of  the Pauli principle violation in the RPA ground state by the calculation of molecular dipole moments and electronic kinetic energies.
A dipole moment is one the best observable for our purposes since it is directly related to the single-particle density matrices and it has readily available accurate experimental values for many molecules. We also computed electronic kinetic energy to assess the role of Pauli principle in the calculations of energy related quantities. The kinetic energy is chosen since it is the only part of electronic energy expressed in terms of expectation values of a single-particle operator.

Let us consider dipole moment operator written in the second quantised form
\begin{equation}
d=   \sum_\sigma \sum_{kl} \langle k | d | l \rangle c^\dag_{k\sigma} c_{l \sigma},
\label{d1}
\end{equation}
where  $ \langle k | d | l \rangle $ is the matrix element of dipole moment between Hartree-Fock orbitals.
We perform the derivations for the dipole moment, the calculations of electronic kinetic energies can be performed along the exactly the same lines.
Transforming it to Hartree-Fock quasiparticle basis (\ref{a+},\ref{a}), we get
\begin{eqnarray}
 d=  2 \sum_{h}  \langle h | d | h \rangle 
+   \sum_{pp'} \langle p | d | p'  \rangle   \rho_{p'p}  -  \sum_{hh'}  \langle h | d | h' \rangle   \rho_{h' h}.
\label{dipole}
\end{eqnarray}
Here the  first term is  the Hartree-Fock value for the molecular dipole moment. We omit in (\ref{dipole}) terms proportional to 
  $a^\dag_{p\sigma} a^\dag_{ h \sigma}$  and 
$a_{h\sigma} a_{ p \sigma}$. These terms
are linearly proportional to  $C^\dag_{ph}(JM)$ and $C_{ph}(JM)$,  respectively, therefore they are also linear in excitation creation $Q^\dag_{JMi}$ and annihilation operators $Q_{JMi}$ and do not contribute to RPA ground state expectation value  due to (\ref{vacuum}). 

Using the expressions obtained in the previous section for single particle density matrices in the RPA ground state, we obtain the following equation for the molecular dipole moment
\begin{eqnarray}
&& \langle \Psi_0 | d | \Psi_0 \rangle =  2 \sum_{h}  \langle h | d | h \rangle 
\label{d}
\\
&&+   \frac{1}{2} \sum_{i} \sum_{JM}  \left(\sum_{hpp'} \langle p | d | p'  \rangle  Y^{Ji}_{p' h}  Y^{Ji}_{p h} 
  -  \sum_{hph'}  \langle h | d | h' \rangle   Y^{Ji}_{p h'}  Y^{Ji}_{p h}\right).
\nonumber
\end{eqnarray}
The extra factor $1/2$ originates from the restoration of the Pauli principle in the RPA ground state.

We compute molecular dipole moments and electronic kinetic energies for a representative set of molecules (lithium hydride, hydrogen fluoride, hydrogen chloride,  water, hydrogen sulfide, ammonia, methyl fluoride, and methanol).
All calculations have been performed with a development version of Mendeleev computer program for {\it ab initio} quantum chemical calculations \cite{mendeleev}. The current RPA implementation uses  Hartree-Fock  reference molecular orbitals and the details of the numerical implementation of RPA method is described in \cite{kosov_rpa17}.
 RPA with the Pauli principle correction calculations  are compared with the standard RPA  approach as well as with the Hartree-Fock (HF),  M{\o}ller-Plesset second order perturbation theory (MP2) and coupled-cluster singles and doubles method (CCSD). We used 6-311++G(3df,3pd) basis set (cartesian form)  in all  calculations. The molecular geometries  are taken from NIST Computational Chemistry Comparison and Benchmark Database \cite{nist}. 
The computed values of molecular dipole moments are also assessed against the experimental results.
\begin{figure}[t!]
\begin{center}
\includegraphics[width=\columnwidth]{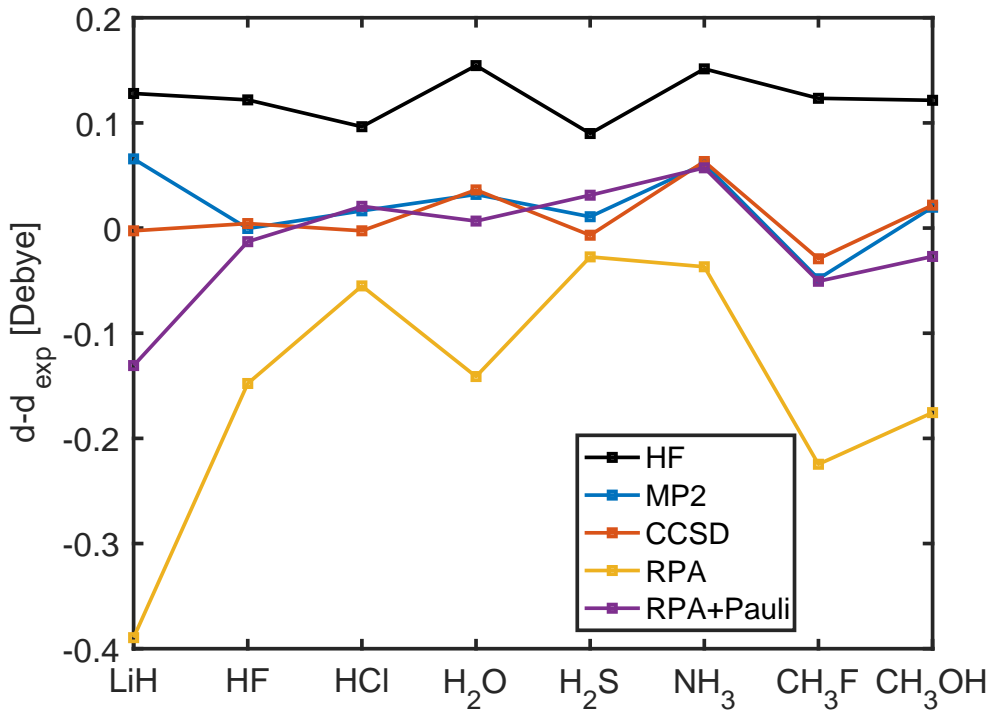}
\end{center}
	\caption{Difference between computed and experimental values of the dipole moments. The experimental values of the dipole moments are taken from NIST Computational Chemistry Comparison and Benchmark Database \cite{nist}.}	
\label{wtd1-fig}
\end{figure}

\begin{figure}[t!]
\begin{center}
\includegraphics[width=\columnwidth]{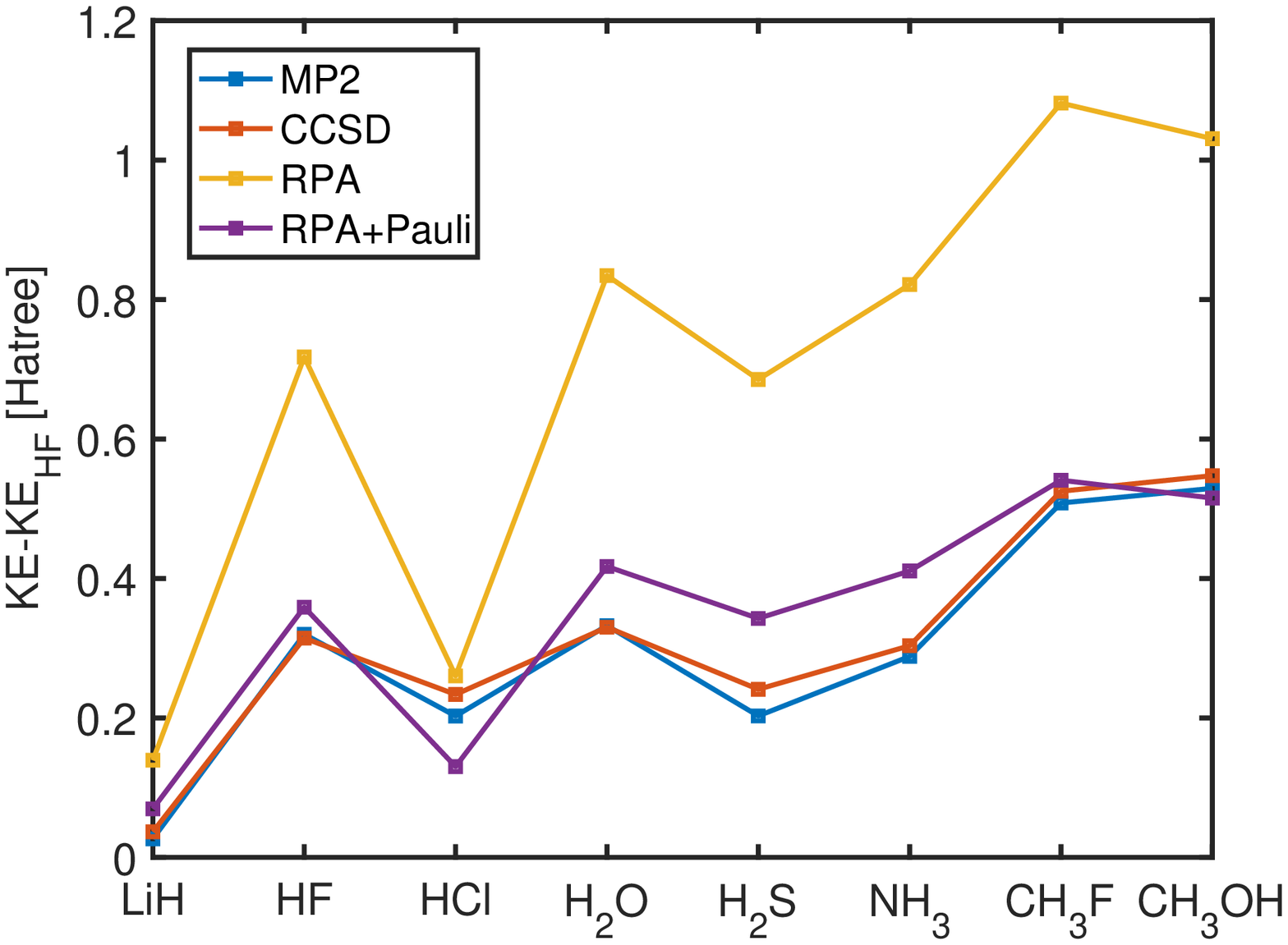}
\end{center}
	\caption{Difference between correlated and Hartree-Fock electronic kinetic energies.}	
\label{wtd1-fig}
\end{figure}

The results of the calculations are summarised in Fig.1 and Fig.2. As seen in Fig.1, Hartree-Fock theory systematically overestimates the molecular dipole moments. The RPA (without correction for  violation of the Pauli principle) produces too strongly correlated electronic ground states lowering  the value of the dipole moment well below the experimental results. Restoring the Pauli principle in the RPA ground state (RPA+Pauli curve in Fig.1) brings the computed values close to the experimental data and gives the accuracy on a par with MP2 and CCSD methods. The similar behaviour is observed for electronic kinetic energies (Fig.2). The standard RPA significantly overestimates the kinetic energy while RPA with Pauli corrections yields the results close to  the MP2 and CCSD values.

\section{Discussions and conclusions}
One of the main results of the paper are equations (\ref{tx})  and (\ref{rhopp-P}). 
Eq.(\ref{tx}) determines the operator $S$ in the RPA ground state $e^S | HF \rangle $ considering the exact commutation relations between particle-hole pair excitation operators (\ref{commutator}), therefore it
takes into account of the Pauli principle in construction of the ground state wavefunction. 
The additional  Pauli correction term in (\ref{tx}) leads to the extra term in the single particle density matrix. The Pauli principle correction to the single particle density matrix is the same density matrix itself taken with negative sign, see
(\ref{rhopp-P}).  The  proper consideration of the Pauli principle leads to  extra $\frac{1}{2}$ factor in the expressions for molecular 
observables (\ref{rhopp-P1},\ref{rhohh-P1},\ref{d}).

To summarise, we have developed  theoretical method to restore the Pauli exclusion principle in the RPA correlated wavefunction. 
We obtained new, corrected for the Pauli principle  expression for the RPA ground state correlated wavefunction.
The method is {\it a posteriori} approach: The RPA eigenvalue problem remains intact and only the final expressions for molecular observables  are corrected by $\frac{1}{2}$ factor.
The role of the  proposed Pauli principle corrections was illustrated by the calculations of molecular dipole moments and electronic kinetic energies. The corrected for the Pauli principle RPA  ground gives results of comparable accuracy with M{\o}ller-Plesset second order perturbation theory and coupled-cluster singles and doubles method, while standard RPA significantly overestimates  the role of electronic correlations. 

 \section*{Acknowledgement}
 
We thank Alan Dzhioev for many valuable discussions.
 
 \appendix
 
 \section{Derivation of eq.(\ref{second})}

  \begin{eqnarray}
&&\sum_{J_1}  \sum_{i} \sum_{JM}
 \sum_{h_1 p_2 h_2 } T^{J_1}_{p' h_2 p_2 h_1} W(J_1,J) X^{Ji}_{p_2 h_2} 
 \langle \Psi_0 | C^\dag_{p h_1} (JM)  
Q^\dag_{\overline{JM} i} |\Psi_0 \rangle 
\nonumber 
\\
&&= \sum_{J_1} \sum_{JM} 
 \sum_{h_1 p_2 h_2 } T^{J_1}_{p' h_2 p_2 h_1} W(J_1,J) 
 \langle \Psi_0 | C^\dag_{p h_1} (JM)  
C^\dag_{p_2 h_2} (\overline{JM})  |\Psi_0 \rangle
\nonumber
\\
&&= \sum_{J_1} \sum_{JM} 
 \sum_{h_1 p_2 h_2 } T^{J_1}_{p' h_2 p_2 h_1} W(J_1,J) 
 \sum_{\sigma_1 \sigma_1' \sigma_2 \sigma_2'} \langle \frac{1}{2} \sigma_1  \frac{1}{2} \sigma_1' | JM \rangle \langle \frac{1}{2} \sigma_2  \frac{1}{2} \sigma_2' | J-M \rangle 
 \nonumber
 \\
 && \times
 (-1)^{J+M}
  \langle \Psi_0 |
  a^\dag_{p\sigma_1} a^\dag_{h_1 \overline{\sigma_1'}} a^\dag_{p_2 \sigma_2} a^\dag_{h_2 \overline{\sigma_2'}}
    |\Psi_0 \rangle 
    \nonumber 
    \\
&&= - \sum_{J_1} \sum_{JM} 
 \sum_{h_1 p_2 h_2 } T^{J_1}_{p' h_2 p_2 h_1} W(J_1,J) 
 \sum_{\sigma_1 \sigma_1' \sigma_2 \sigma_2'} \langle \frac{1}{2} \sigma_1  \frac{1}{2} \sigma_1' | JM \rangle \langle \frac{1}{2} \sigma_2  \frac{1}{2} \sigma_2' | J-M \rangle 
  \nonumber
 \\
 && \times
 (-1)^{J+M}
  \langle \Psi_0 |
  a^\dag_{p\sigma_1} a^\dag_{h_2 \overline{\sigma_2'}} a^\dag_{p_2 \sigma_2} a^\dag_{h_1 \overline{\sigma_1'}}
    |\Psi_0 \rangle 
    \nonumber 
    \\
    &&= - \sum_{J_1} \sum_{JM} 
 \sum_{h_1 p_2 h_2 } T^{J_1}_{p' h_2 p_2 h_1} W(J_1,J) 
 \sum_{\sigma_1 \sigma_1' \sigma_2 \sigma_2'} \langle \frac{1}{2} \sigma_1  \frac{1}{2} \sigma_1' | JM \rangle \langle \frac{1}{2} \sigma_2  \frac{1}{2} \sigma_2' | J-M \rangle 
 \nonumber
 \\
&& 
\times  (-1)^{J+M} \sum_{J_2M_2} \sum_{J_3 M_3} \langle \frac{1}{2} \sigma_1  \frac{1}{2} \sigma_2' | J_2 M_2 \rangle \langle \frac{1}{2} \sigma_2  \frac{1}{2} \sigma_1' | J_3 -M_3 \rangle (-1)^{J_3 +M_3}
 \nonumber
 \\
&& 
\times
\underbrace{ \langle \Psi_0 | C^\dag_{p h_2} (J_2 M_2)  
C^\dag_{p_2 h_1} (\overline{J_3 M_3})  |\Psi_0 \rangle}_{\sim \delta_{J_2 J_3} \delta_{M_2 M_3} }
    \nonumber 
    \\
    &&=
    -
    \sum_{J_1} \sum_{J_2 M_2} \sum_{h_1 p_2 h_2} T^{J_1}_{p'  h_1 p_2 h_2} \; \underbrace{ \sum_{J} W(J_1,J) W(J, J_2)}_{\delta_{J_1 J_2}} 
     \nonumber
 \\
&& 
\times
    \langle \Psi_0 | C^\dag_{p h_1} (J_2 M_2)  
C^\dag_{p_2 h_2} (\overline{J_2 M_2})  |\Psi_0 \rangle
    \nonumber 
    \\
    &&=
       -  \sum_{J_2 M_2} \sum_{h_1 p_2 h_2} T^{J_1}_{p'  h_1 p_2 h_2} 
    \langle \Psi_0 | C^\dag_{p h_1} (J_2 M_2)  
C^\dag_{p_2 h_2} (\overline{J_2 M_2})  |\Psi_0 \rangle = - \langle \Psi_0 | \rho_{p'p} |\Psi_0 \rangle
\nonumber
\end{eqnarray}

 \newpage
 \section*{References}

\end{document}